\documentclass[reprint, amsmath,amssymb, aps, prl]{revtex4-1}

\usepackage{epsfig}
\usepackage{graphicx}
\usepackage{dcolumn}
\usepackage{bm}

\usepackage[usenames,dvipsnames]{xcolor}

\newcommand{\ket}[1]{\ensuremath{\left|#1\right\rangle}}

\begin{document}

\title{A proposed test of quantum mechanics with three connected atomic clock transitions}


\author{Mark G. Raizen}
\affiliation{Department of Physics, The University of Texas at Austin, Austin, Texas, 78712}

\author{Gerald Gilbert}
\affiliation{MITRE-Princeton, 200 Forrestal Road, Princeton, NJ 08540}

\author{Dmitry Budker} 

\affiliation{Johannes Gutenberg-Universit{\"a}t Mainz, Helmholtz-Institut Mainz, GSI Helmholtzzentrum f{\"u}r Schwerionenforschung, 55128 Mainz, Germany}

\affiliation{Department of Physics, University of California, Berkeley, California 94720, USA}

\date{\today}

\begin{abstract}

We consider possible extensions to quantum mechanics proposed by Steven Weinberg, and re-analyze his prediction of a new test based upon three atomic clocks in the same atom. We propose 
realistic experimental systems where this hypothesis can be tested.  Two systems already set limits on deviations from quantum mechanics, while with another system, one would be able to search for new physics at the limit of sensitivity of the best atomic clocks.

\end{abstract}

\pacs{Valid PACS appear here}

\maketitle
\noindent

\section{Introduction}

\begin{figure}[!htpb]\centering
    \includegraphics[width=\linewidth]{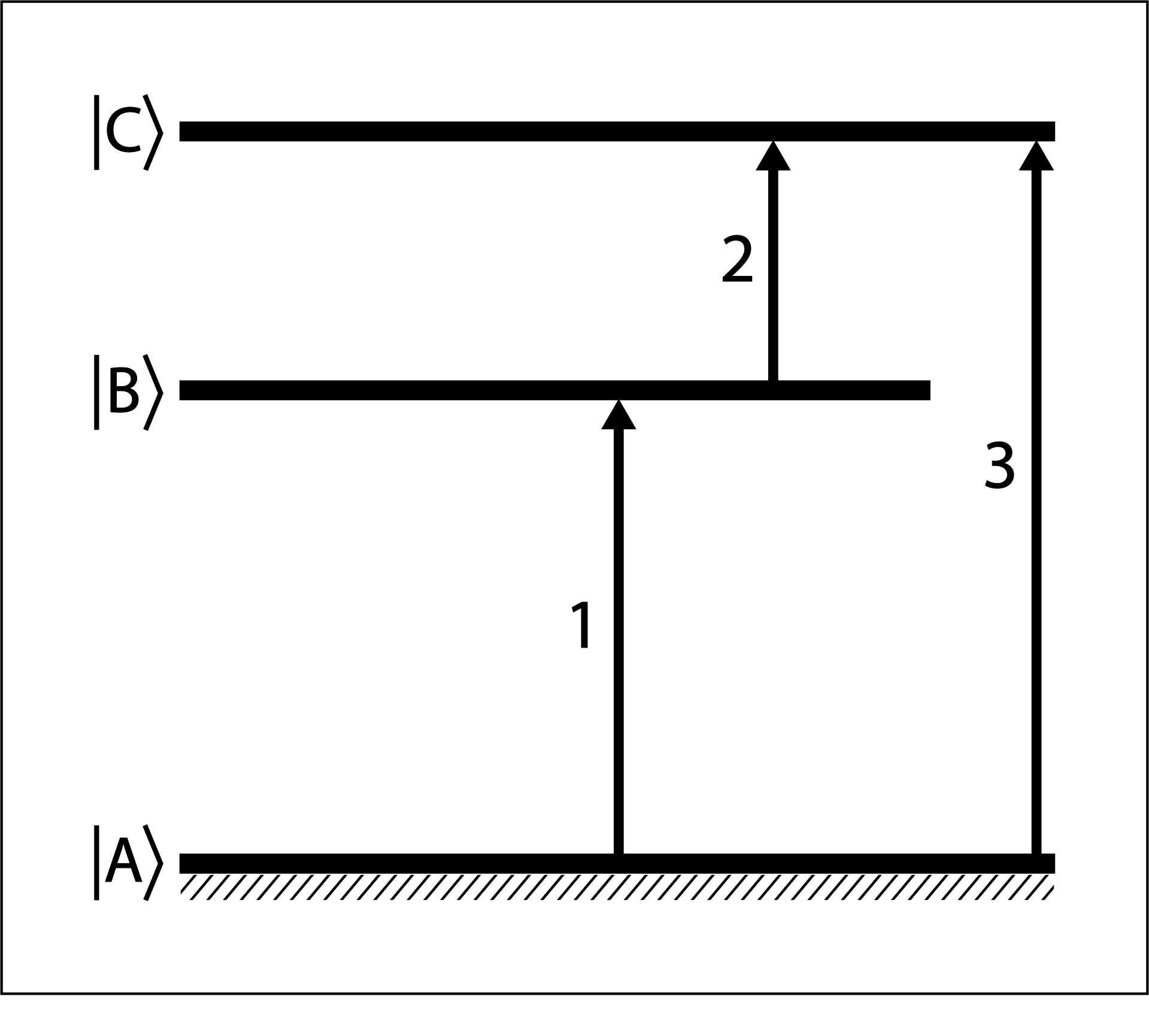}
    \caption{The general idea of testing for a possible nonlinearity effect as proposed by Steven Weinberg. 
    }
    \label{Fig:123}
\end{figure}

Quantum mechanics is one of the fundamental pillars of modern physics. It has remained essentially unchanged since it was developed almost 100 years ago \cite{Weinberg2015lectures}. Although in continuous practical use since its discovery, there is general agreement that quantum mechanics is an incomplete theory as it does not incorporate interactions with the environment which are mostly unavoidable.  For example, even atoms in isolation in a laboratory vacuum are immersed in black-body radiation or other external electromagnetic fields. The Schr\"{o}dinger equation is time-symmetric, with no past and no future, just state vectors, eigenvalues, and Hermitian operators. This creates a tension with what is referred to as the macroscopic world, where there is a clear arrow of time. Moreover, the underlying, dynamical assumptions of quantum mechanics do not specify precisely how to account for a measurement apparatus, requiring additional prescription from outside those assumptions. In an attempt to resolve the situation, the concept of measurement-induced collapse of the wave-function was invoked, known as the Copenhagen interpretation. An alternative approach is the many-worlds interpretation, where endless branches of parallel histories are created. Neither are appealing options, and do not fit into a consistent framework that extends quantum mechanics.  There are also no experimental tests that can point to one or the other, a very unsatisfactory state of affairs.  This led Steven Weinberg to propose in 2014 that quantum mechanics can be formulated without the need for state vectors \cite{Weinberg2014QMwithout}. In 2016 Weinberg further proposed the possibility of observing small departures from ordinary quantum mechanics by studying implications of the Lindblad equation that might be detected by utilizing the great precision of atomic clocks \cite{Weinberg2016Lindblad}. Weinberg proposed that this would have experimentally measurable consequences, specifically with atomic clocks which have now reached an absolute accuracy of a part in 10$^{18}$.  He envisioned an atom with three clock transitions, where each one could be measured independently of the other two.  These transitions would be from state $\ket{A}$ to $\ket{B}$ denoted 1, $\ket{B}$ to $\ket{C}$ denoted 2, and $\ket{A}$ to $\ket{C}$, denoted 3, illustrated schematically in Fig.\,\ref{Fig:123}. The manifestation of the extension of quantum mechanics would be that the sum of the first two clock frequencies would not be equal to the third, i.e., 1+2\,$\ne$\,3. Weinberg did not suggest in \cite{Weinberg2016Lindblad} any concrete examples where this hypothesis could be tested.  

In this paper, we make the case for 
experimental systems to search for new physics beyond standard quantum mechanics.  

Two systems are a trapped $^{226}$Ra ion, and a trapped $^{40}$Ca ion where three such transitions have, in fact, been measured, though they are not all atomic clock transitions thereby precluding high accuracy.  The newly proposed system is neutral $^{171}$Yb atoms which offer the potential for extremely high accuracy typical of the best atomic clocks.  

While Weinberg's extension of quantum mechanics has been a hitherto untested hypothesis, his theory does not provide an exact quantitative prediction of the size of the effect. Thus, tests in different systems are particularly useful for probing the idea.


 
\section{Existing limits}

\begin{figure}[!htpb]\centering
    \includegraphics[width=\linewidth]{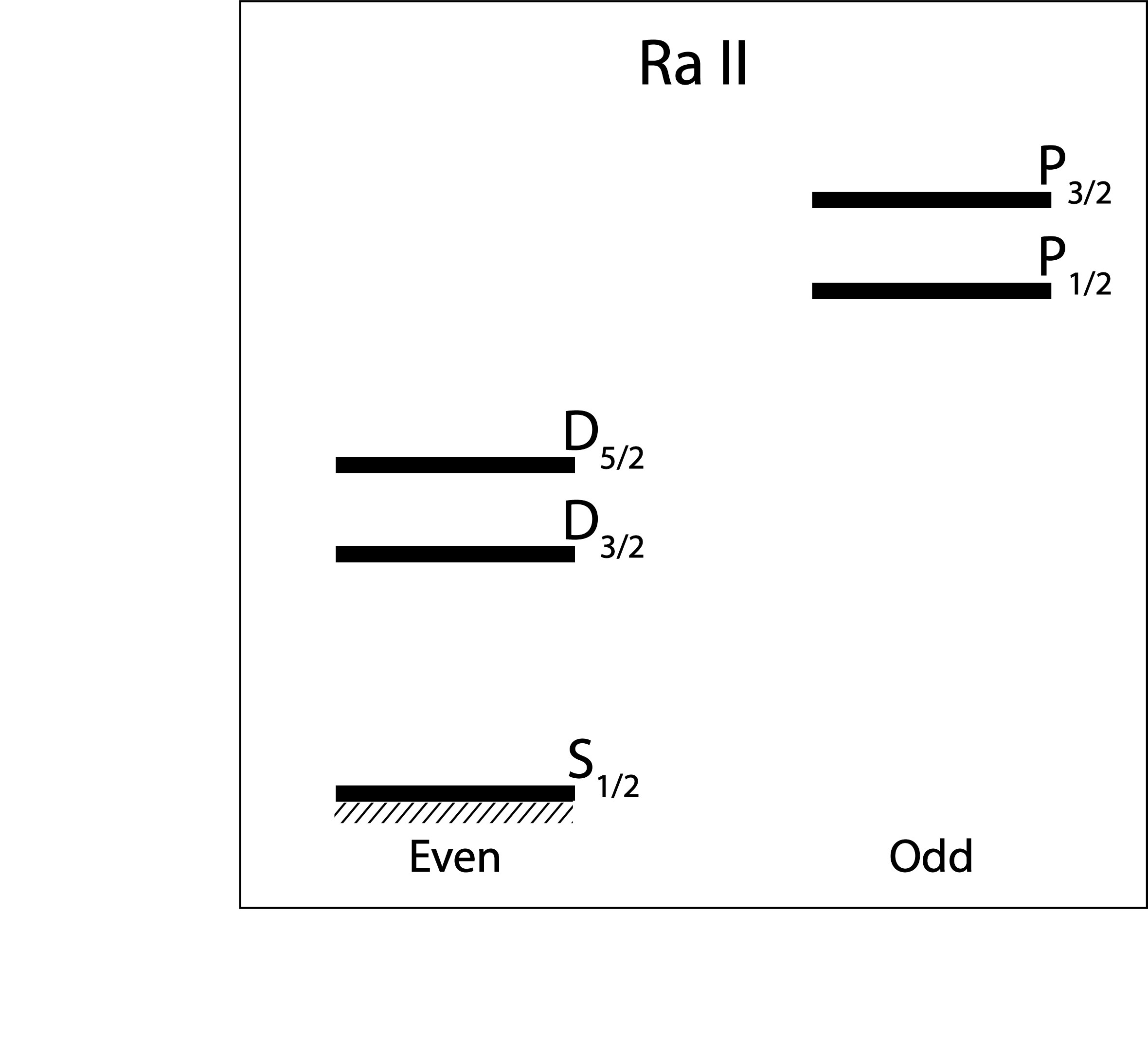}
    \caption{Selected low-lying energy levels of Ra II \cite{NIST_ASD} (not to scale).
    }
    \label{Fig:226Ra_levels}
\end{figure}

In searching for possible experimental systems where there are three transitions that are tied together, we found existing data for a single trapped $^{226}$Ra ion. While radium does not have any stable isotopes, the half-life of $^{226}$Ra is about 1585.5\,y. It is a daughter in the decay chain of $^{238}$U, so can be found in nature as first discovered by Marie and Pierre Curie in 1898. Singly-ionized radium was confined in a RF Paul trap and spectroscopic measurements were reported in a series of papers \cite{Holliman2019_226Ra,Holliman2020_Direct,Holliman_2022_Ra}. These papers did not cite the 2016 Weinberg paper, so, evidently, the authors were not aware of that work.  A schematic of the energy levels is shown in Fig.\,\ref{Fig:226Ra_levels}. The ground state is $7s~^2S_{1/2}$, and two electric quadrupole transitions, $7s~^2S_{1/2}\rightarrow 6d~^2D_{5/2}$ at 728\,nm and $7s~^2S_{1/2}\rightarrow 6d~^2D_{3/2}$ at 828\,nm. In addition, there are two electric dipole transitions $6d~^2D_{3/2}\rightarrow 7p~^2P^o_{3/2}$ at 708\,nm and $6d~^2D_{5/2}\rightarrow 7p~^2P^o_{3/2}$ at 802\,nm. In a subsequent paper, the electric dipole transition $7s~^2S_{1/2}\rightarrow 7p~^2P^o_{3/2}$ at 382\,nm was reported. To within the accuracy of these measurements, the results do not exhibit any additional frequency shifts, but the precision was limited by the fact that only one of the three transitions was an electric quadrupole, while the other two were electric dipole.  The quoted accuracy is around 30\,MHz, though could presumably be improved considerably with longer interrogation times.  Nevertheless, the $^{226}$Ra system does not satisfy the criterion proposed by Weinberg of having three connected atomic clocks.  The same data provide other combinations of transitions, for example 828\,nm and 708\,nm reach the same level as 728 and 802\,nm. Again, only two of these transitions (828\,nm and 728\,nm) are electric quadrupole, while the other two are electric dipole. 

A second case where there are existing data is a single trapped $^{40}$Ca ion (Ca\,II).  The atomic structure is similar to that of the radium ion, though the transition frequencies and lifetimes are different. Transitions in Ca\,II were investigated in the context of precision determination of isotope shifts \cite{Kramida2020_Ca} and ``transition closures'' were studied at the 100\,kHz level \cite{Mueller2020_Collinear}, although precise calculation (rather than direct measurement) was used for some of the transitions.
A fully experimental test is possible using the electric quadrupole transition,  $4s~^2S_{1/2}\rightarrow 3d~^2D_{5/2}$ at 729\,nm, which was accurately measured \cite{Huang2016_CaII} combining them with the electric dipole transitions $4s~^2S_{1/2}\rightarrow 4p~^2P^o_{3/2}$ at 393\,nm and $4s~^2D_{5/2}\rightarrow 4p~^2P_{3/2}$ at 854\,nm
reported \cite{Mueller2020_Collinear}.  Currently, the accuracy of the limits in Ca\,II is around 100\,kHz, considerably better than in Ra\,II, and also may be improved in the future.

We now propose another atomic system where there are three connected clock transitions.
 
\section{Proposed Test: 1-2-3 $^{171}$Yb ensemble clocks}

\begin{figure}[!htpb]\centering
    \includegraphics[width=\linewidth]{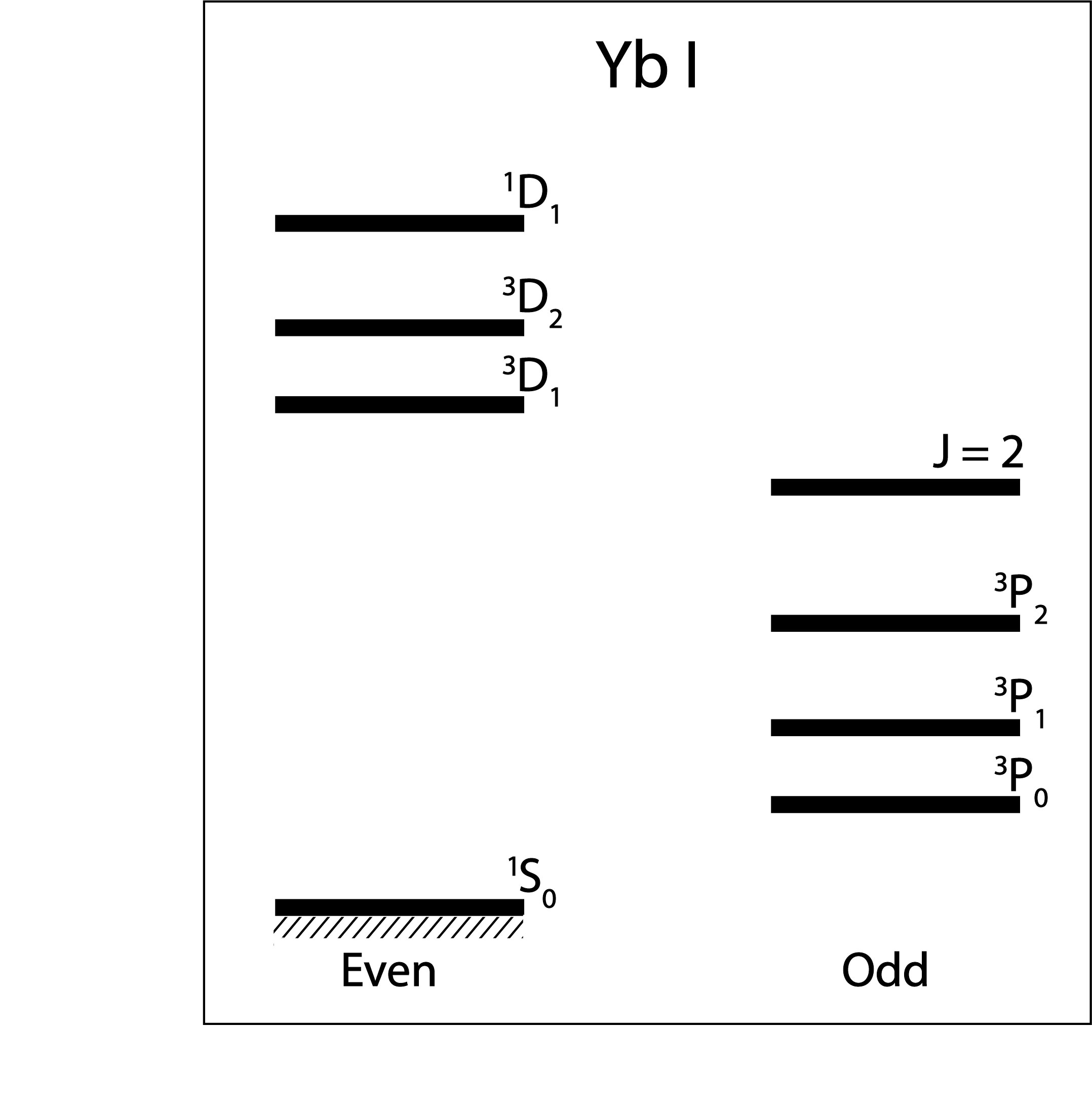}
    \caption{Selected low-lying energy levels in Yb\,I \cite{NIST_ASD} (not to scale). Their properties relevant to the 1-2-3 test are listed it Table\,\ref{Tab:Yb_states}
    }
    \label{FIG:Yb_levels}
\end{figure}

Various options for testing for Weinberg's nonlinearity are offered by atomic system with two electrons above closed shells, for example, neutral ytterbium (see Fig.\,\ref{FIG:Yb_levels}). Such atoms have an additional experimental advantage in that there are well-developed techniques for their trapping and cooling and that lattice clocks based on such atoms have been demonstrated \cite{Ludlow2015RMP_Clocks}, The natural choice of state 1 is the ground 4f$^{14}$6s$^2$ state, while state 3 could be chosen from one of the $J=1,2$ states of the same (even) parity. The even-parity $J=2$ states are connected to the ground state via electric quadrupole (E2) transitions as well as by degenerate two-photon transitions (i.e., transitions that can be driven by photons of the same wavelength). The $^1S_0\rightarrow$ $^1D_2$ transition was recently used for a precision (uncertainty of $\approx 300$\,Hz) determination of isotope shifts \cite{Figueroa2021_Yb_IS}.  As for the transitions between the ground state and $J=1$ even-parity states, they are forbidden by E2 selection rules, as well as by selection rules for degenerate two-photon transitions, The way to drive these transitions is to use nondegenerate photons or, alternatively, to rely on mixing of $J$ states by hyperfine interactions or magnetic fields, see \cite{BowersPRA1996,Kozlov2009_BEsuppressed_2Phot} and references therein. In order to ultimately achieve high spectroscopic resolution in the 1-3 transition, it is prudent to choose the upper level with the longest possible radiative lifetime. The 4f$^{14}$5d6s\,$^1$D$_2$ state used for the the isotope-shift measurement has a lifetime of $\approx 6.7\,\mu$s \cite{BowersPRA1996}. The E2 transition from the ground state is at 361\,nm, while the degenerate two-photon transition occurs at twice that wavelength (i.e., around 723\,nm). 
\begin{table}[htpb]
    \centering
    \begin{tabular*}{\linewidth}{@{\extracolsep{\fill}} lccc}
    \hline 
    \hline
 State  & & Energy (cm$^{-1}$) \cite{NIST_ASD}   & Lifetime \\
    \hline
    4f$^{14}$6s$^2$ & $^1$S$_0$ & 0 &   \\
    \hline
    4f$^{14}$6s6p & $^3$P$_0$ & 17\,288.439  & $\approx$20\,s\,\cite{Safronova2018_Two_Yb_clocks}  \\
     & $^3$P$_1$ & 17\,992.007  & 866\,ns\,\cite{Beloy2012Yb}  \\
     & $^3$P$_2$ & 19\,710.388  & $\approx$9\,s\,\cite{MISHRA2001_Rad}  \\
    \hline
    4f$^{13}$5d6s$^2$ & J=2 & 23\,188.518   & $\approx$1\,min\,\cite{Dzuba2018Yb_clock,Safronova2018_Two_Yb_clocks}  \\
    \hline
    4f$^{14}$5d6s & $^3$D$_1$ &   24\,489.102     & 329\,ns\,\cite{Beloy2012Yb}  \\
     & $^3$D$_2$ &   24\,751.948     & 460\,ns\,\cite{BowersPRA1996}  \\
  & $^1$D$_2$ &   27\,677.665      & 6.7\,$\mu$s\,\cite{BowersPRA1996}  \\
     
    \hline
    \hline
    \end{tabular*}
    \caption{Properties of Yb\,I states shown in Fig.\,\ref{FIG:Yb_levels}. Lifetimes of metastable states depend on hyperfine mixing and are given for $^{171}$Yb. }
    \label{Tab:Yb_states}
\end{table}

We now discuss the choice of level 2 and the 1-2 and 2-3 transitions. One option for level 2 is the metastable 4f$^{14}$6s6p\,$^3$P$_0$ state. Here, the 1-2 transition can occur due to hyperfine mixing with the $^3P_1$ state for an isotope with nonzero nuclear spin (e.g., $^{171}$Yb with $I=1/2$. The advantage here is that existing ytterbium lattice clocks already operate on this transition. The 2-3 transition can proceed by a similar mechanism and the clock operation on several transitions in the same atom can proceed as proposed in \cite{Safronova2018_Two_Yb_clocks} by alternating measurements on different transitions, with adiabatic switching of the lattice wavelength to avoid systematics. 

For an even narrower 1-3 transition, one may choose to work with the metastable odd-parity $J=2$ state at 23188.518\,cm$^{-1}$, the lifetime of which is predicted to be on the order of 1\,min \cite{Dzuba2018Yb_clock,Safronova2018_Two_Yb_clocks}. In fact, this is exactly the system considered in \cite{Dzuba2018Yb_clock,Safronova2018_Two_Yb_clocks} for the dual atomic clock in one and the same atom in the context of the search for possible variation of fundamental constants \cite{Dzuba2018Yb_clock,Safronova2018_Two_Yb_clocks}. For the 1-3 transition, one would use either an E1-M1 two-photon transition or, alternatively, a hyperfine-interaction induced two-photon transition at 852\,nm. 

Several specific schemes exist in Yb for the 1-2-3 test, and it appears that such tests can be carried out with existing technologies, easily to sub-Hz accuracies in the initial experiments with lattice clocks, and eventually down to the full capacity of such clocks, i.e., another several orders of magnitude better.   

\section{Outlook and Conclusions}

The cases considered here are representative; other combinations of clock transitions are possible. Interesting possibilities for 1-2-3 test exist with molecules, for instance, the HD$^+$ molecular ions \cite{Kortunov2021_HD+}, transitions in highly charged relativistic ions \cite{Wojtsekhowski2021_LLI}, and perhaps even with nuclear gamma transitions \cite{Budker2022_GF_Nuclear}. 

The highest accuracy with neutral atoms to date has been reported with Yb and Sr atoms trapped in optical lattices, and other promising clock transitions are currently being investigated.  Three clock transitions in an atom are the minimum number for testing Weinberg's hypothesis, though more transitions could provide cross checks and higher sensitivity \footnote{We note that four-transition closures were investigated via collinear spectroscopy of Ca$^+$ \cite{Mueller2020_Collinear}.}. 

Once the 1-2-3 tests start pushing the limits of sensitivity, one will need to contend with subtle effects that determine the generally asymmetric lineshapes observed in experiment, where the relation between the lineshape parameters and the ``clock readout" will need to be carefully understood. One such effect is quantum interference \cite{Udem2019_QuantInt}, the treatment of which with Lindblad formalism could perhaps be extended to include the Weinberg's extension of quantum mechanics.

In summary, we outline here realistic experimental tests of a possible extension of quantum mechanics first proposed by Steven Weinberg in 2016. If a violation is found, this would require a reformulation of quantum mechanics to account for inevitable coupling to the environment.

\section*{Acknowledgements}
This paper is dedicated to the memory of our mentor and friend, Steven Weinberg. 
We are grateful to Surjeet Rajendran, Nataniel L. Figueroa, Stephan Schiller, Piet Schmidt, Thomas Udem and David Allcock for useful discussions.  The work of MGR was supported by the Sid W. Richardson Foundation. The work of GG was supported by the MITRE Quantum Moonshot program. The work of DB was supported by the Deutsche Forschungsgemeinschaft (DFG) - Project ID 423116110 and by the Cluster of Excellence Precision Physics, Fundamental Interactions, and Structure of Matter (PRISMA+ EXC 2118/1) funded
by the DFG within the German Excellence Strategy (Project ID 39083149). 

\bibliography{bibliography}

\end{document}